# Tunable and broadband coherent perfect absorbers with nonlinear and amplification performance based on asymmetric bifacial graphene metasurfaces

Tianjing Guo and Christos Argyropoulos*

Department of Electrical and Computer Engineering, University of Nebraska-Lincoln,
Lincoln, Nebraska 68588, USA

*christos.argyropoulos@unl.edu

**Abstract:** We present an asymmetric bifacial graphene metasurface to realize tunable and broadband coherent perfect absorption (CPA) at terahertz (THz) frequencies. The proposed design is composed of two dissimilar graphene square patch metasurfaces stacked over an ultrathin dielectric substrate. Nonlinear CPA performance without resorting to optical nonlinearities and amplification are demonstrated based on wave interference phenomena taking place along and inside this subwavelength structure. Broadband asymmetric reflection from opposite directions is obtained in the case of single wave illumination. This response acts as a precursor to the presented tunable and broadband nonlinear CPA performance and the inverse response to absorption, i.e., amplification, when an additional input wave coherently interacts with the opposite propagating incident wave. The proposed ultrathin broadband CPA device is expected to be a vital component to several new on-chip THz communication devices, such as optical modulators, sensors, photodetectors, and all-optical small-signal amplifiers.

## 1. Introduction

Graphene, a two-dimensional (2D) conductive material [1], has attracted considerable attention for its remarkable optical properties and promising applications in the field of photonics and optoelectronics [2–4]. A single graphene monolayer has an extremely low absorptivity of only 2.3%, which results from its sheet conductivity values [2,5,6]. Many works have been reported to increase its absorption by enhancing the light-graphene interaction, such as patterning graphene into ribbons, disks or rectangular patches, adopting a grating coupler, and combining graphene with dielectric layers [7–15]. It is noteworthy that the excitation of graphene plasmons has become an important platform to achieve enhanced light-matter interactions [11,16–19]. Highly confined and tunable surface plasmons at terahertz (THz) frequencies are formed with graphene metasurfaces, making graphene an attractive alternative to traditional noble metal plasmonics. Recently, many exciting applications based on graphene metasurfaces have been realized, such as phase shifters, cloaks, sensors, and polarizers [20–23]. An additional advantageous feature of graphene-based devices compared to metal-based conventional plasmonic structures [24,25] is their wide frequency tunability range, since graphene's Fermi level can be modulated usually by applying a voltage through electrostatic doping or by chemical processing [26,27].

On a relevant context, the phenomenon of coherent perfect absorption (CPA) has recently been proposed and experimentally realized as the time-reversed counterpart of laser [28–32]. However, it has been demonstrated that a narrowband response or single-frequency static (non-tunable) operation is the major limitation in the CPA response [6,32–34], since it is the time-reversed process of laser. By decreasing the size of the absorbing body or increasing its losses, this bandwidth limitation can be relaxed [35,36]. Fortunately, 2D materials can provide a practical alternative to the design of ultrathin CPA devices with broader operation bandwidth. CPA systems based on graphene metasurfaces achieve enhanced incident light absorption and broader CPA bandwidth [37–41]. One more notable feature is the tunability of graphene metasurface-based CPA devices by dynamically manipulating the doping level of graphene. Interestingly, we will present that asymmetric graphene metasurfaces can further increase the CPA operation bandwidth, due to the broadband asymmetric reflection from both sides of the proposed ultrathin absorbing body.

In this work, we demonstrate tunable and broadband CPA response by using a bifacial graphene metasurface illuminated by two counter-propagating incident waves. The proposed metasurface design is asymmetric, which is proven that further facilitates the broadband CPA response. We first derive general CPA conditions for asymmetric structures by using the scattering matrix formalism. Exact formulas for the amplitude ratio and phase difference of the two incident waves are derived to achieve CPA, by computing the reflection and transmission coefficients under single wave illumination from each side. Tunable CPA performance is realized by manipulating the Fermi level of the graphene patches.

The proposed configuration is a four-port device, consisting of two inputs and two outputs. This feature can facilitate a nonlinear optical absorption response and also make the structure to operate as an all-optical passive amplifier [42–45]. The presented structural asymmetry is the key characteristic to enhance these interesting phenomena, bringing them in the extremely compact subwavelength scale, and making them accessible with realistic excitation configurations. The proposed bifacial metasurface design indeed demonstrates tunable and broadband nonlinear CPA response without resorting to the nonlinear optical properties of the used materials. It also acts as an efficient small-signal amplifier without the use of any active (gain) materials. Interestingly, the presented strong amplification effect is the inverse response compared to absorption. The asymmetric bifacial graphene metasurface system can amplify the signal over a broad frequency range in an efficient way with the addition of a tunable performance, which are advantageous features compared to previous works [44]. Both nonlinear and amplifying phenomena are based on wave interference effects along and inside the proposed subwavelength metasurface and not in the materials composing the presented structures. The presented work is expected to facilitate the design of several new compact and planar devices operating at THz frequencies, including modulators, sensors, detectors, and all-optical small-signal amplifiers.

## 2. Theory

The proposed broadband CPA design is composed of two graphene square patch arrays acting as THz resonators stacked over a subwavelength thick dielectric layer, as shown in Fig. 1. Two counter-propagating waves are used to illuminate the bifacial metasurface. The period of both metasurfaces is $P$. The lengths of the graphene square patches in the back and front sides are $L_1$ and $L_2$, respectively. The electric fields $E_{out1}$ and $E_{out2}$ of the outgoing waves are linked to the incident waves $E_{in1}$ and $E_{in2}$ through the scattering matrix $S$ via the formula:

$$\begin{bmatrix} E_{out1} \\ E_{out2} \end{bmatrix} = S \begin{bmatrix} E_{in1} \\ E_{in2} \end{bmatrix} = \begin{bmatrix} r_1 & t_1 \\ t_2 & r_2 \end{bmatrix} \begin{bmatrix} E_{in1} \\ E_{in2} \end{bmatrix} \quad (1)$$

where $r_1, r_2$ and $t_1, t_2$ are the reflection and transmission coefficients when the metasurface is excited by the back or front sides with incident waves $E_{in1}$ and $E_{in2}$, respectively. Note that the transmission coefficient is equal from both incident directions $(t = t_1 = t_2)$ because reciprocity is not broken by this configuration. The relationship between the two input waves is expressed by using their ratio given by the complex variable: $\alpha = E_{in1}/E_{in2}$. As a result, the electric fields $E_{out1}$ and $E_{out2}$ of the outgoing waves become functions of just one incident wave:

$$E_{out1} = (\alpha r_1 + t)E_{in2}$$
$$E_{out2} = (t\alpha + r_2)E_{in2} \tag{2}$$

The CPA operation implies that the outgoing waves need to be completely vanished, i.e., $E_{out1} = E_{out2} = 0$. In order to quantitatively investigate CPA, we define the output coefficient $\Theta$ as the ratio of the total intensity of both output waves to that of the input waves [46,47]:

$$\Theta = \frac{|E_{out1}|^2 + |E_{out2}|^2}{|E_{in1}|^2 + |E_{in2}|^2} = \frac{|t + r_1\alpha|^2 + |t\alpha + r_2|^2}{(1+|\alpha|^2)} \tag{3}$$

CPA is achieved when the output coefficient $\Theta$ becomes zero. The conditions to achieve CPA are derived by Eq. (3) to be:

$$\alpha = -t/r_1 \text{ and } \alpha = -r_2/t \tag{4}$$

where $\alpha$ is the ratio of the input waves. This complex variable can be rewritten as: $\alpha = |\alpha|e^{i\varphi}$, where $\varphi$ is the phase difference between $E_{in1}$ and $E_{in2}$. Thus, in order to realize CPA, we need to optimize the proposed structure to get specific reflection and transmission coefficients from each side illumination and meet the CPA conditions given by Eq. (4). Note that the asymmetry in reflection arises from the two different metasurfaces used in the bifacial design and will be utilized to facilitate the achievement of the tunable, broadband and nonlinear CPA response and the presented amplifying performance.

## 3. Results and Discussion

The graphene is sufficiently thin compared to the wavelength used in this work. Thus, it is modelled as an ultrathin sheet with complex surface conductivity $\sigma_g$ that can be tuned by chemical doping or external electric field bias [48,49]. A surface current is excited along each graphene patch by the incident waves. The conductivity of graphene is commonly described by the Kubo formula [50], consisting of both interband and intraband electron transitions. In this work, we ignore the interband transitions because the device is designed to operate at THz frequencies below the interband transition threshold, leading to blocked interband transitions [7,51]. The complex surface conductivity, arising from the intraband electron transition, is well described by using a Drude expression [52]: $\sigma_g = -ie^2 E_F/(\pi\hbar^2(\omega - i2\Gamma))$, where $e$ is the electron charge, $E_F$ is the doping level or Fermi level, $\hbar$ is the reduced Plank constant, $\omega = 2\pi f$ is the angular frequency, and $\Gamma$ is the phenomenological scattering rate, given by $\Gamma = ev_F^2/\mu E_F$, where $v_F \approx c/300$ is the Fermi velocity and $\mu = 1 m^2/Vs$ is the measured DC mobility, consisting typical values for graphene [53]. These formulas are valid for the currently used room temperature simulations.

In the proposed design, the monolayer graphene is patterned into a periodic array of patch resonators with different dimensions at each side. The period of the square unit cell is *P= 7.6um*, the side lengths of the graphene square patches on the back or front side are *L₁=7.4um* and *L₂=5um*, respectively. The thickness of the dielectric layer between the two asymmetric graphene patch arrays is *t=2um* and its dielectric permittivity is $\varepsilon = 2.25$, similar to silica. The dielectric material used is lossless but improved CPA performance will be achieved in case a lossy dielectric material will be utilized in the proposed metasurface design. Thinner dielectric layers can also be used but this will reduce the total asymmetry of the proposed structure, thus making the obtained CPA response more narrowband. Thicker dielectric layers will not significantly affect the presented broadband CPA response, as well as larger periodicity values. The Fermi level of the graphene monolayer is fixed to 0.4 eV. We will keep using the same parameters and normal incident angle illumination scenario during this work unless otherwise specified. The schematic of the proposed design is shown in Fig. 1. COMSOL Multiphysics [54] is employed to optimize this design and to numerically simulate the transmission, reflection, and absorption coefficients under single wave illumination and the coherent absorption performance under two counter-propagating incident waves. Periodic boundary conditions are utilized in the x- and y- direction, while port boundaries are placed in the z-direction to create the counter-propagating incident waves. In the case of a potential experimental characterization, the proposed design can be fabricated by using conventional fabrication techniques, such as chemical vapor deposition and electron-beam lithography [55,56]. In addition, ion gel with low relative permittivity of 1.82 can be used to serve as the gate dielectric material on top of the graphene metasurfaces. Two gold gate contacts can be fabricated on the ion gel layers [27,57–59], as shown in the inset of Fig. 1, to apply voltage, dope the graphene, and, as a result, tune the bifacial metasurface response. Due to the thin thickness and low permittivity of the ion gel layers, their effect on the performance of the system was found to be negligible (not shown here) and is ignored in our simulations.

The computed reflection, transmission and absorption coefficient spectra of the presented structure under a single incident transverse magnetic (TM)-polarized wave at normal incidence are shown in Fig. 2 for back (a) and front (b) side illumination. Note that the presented structure will also exhibit identical response to transverse electric (TE)-polarized illumination, since square patches are used. Hence, the presented metasurface can work for both polarizations, a major advantage for several potential applications [20–23]. Close to 50% absorption is achieved in a broad frequency range for both directions of the single incident wave, as it is shown in Fig. 2. The 50% absorption value limit demonstrates critical coupling. This is the maximum absorption that can be attained by an ultrathin symmetric subwavelength structure [60]. Note that thanks to the currently proposed extremely asymmetric design, the absorption in the case of front side illumination breaks the 50% absorption bound [10]. This effect lays the groundwork to realize the currently presented broadband CPA response. The transmission and reflectioncoefficients for back side illumination are $T_1 = 25.9\%$ and are $T_2 = 25.9\%$ and $R_2 = 21.4\%$ for front side illumination at the same frequency. The transmission from both sides is identical because reciprocity is not broken with this configuration. However, reflection is different due to the structural asymmetry, leading to the presented nonlinear CPA response.

Next, we launch two counter-propagating TM-polarized incident waves from the front and back sides of the bifacial metasurface. We vary the phase difference $\varphi$ between the two incident waves and compute the metasurface CPA response. The amplitude $|\alpha|$, which is equal to the ratio of $E_{in1}$ over $E_{in2}$, is chosen to be to be approximately 0.9, to fulfill the CPA conditions described by Eq. (4). The computed output coefficient, as a function of frequency and phase difference between the two input waves, is shown in Fig. 3(a). Nearly perfect absorption is obtained over a broad frequency range by using a wide phase difference $\varphi$ range. Notable, the output coefficient is modulated from zero at the CPA frequency of 2.4 THz to 99.6% by adjusting the phase difference between the two incident waves, as seen in Fig. 3(a). Hence, the input signals are modulated from being completely absorbed by the proposed ultrathin device to passing almost unaffected (completely transparent device). This result can lead to several potential applications, such as planar THz modulators or graphene-based coherent photodetectors [61,62]. The modulation depth [28], defined as the ratio of the maximum output coefficient to the minimum one, can reach values as high as 5000, which is impressive, given that the thickness of the proposed device is 1/60 of the operating wavelength. The phase difference value of the two incident waves is fixed to $\varphi = 10°$ to calculate the corresponding output coefficient only as a function of frequency. This particular phase value is chosen because it corresponds to the best absorption performance, where destructive interference between the two incident waves occurs. The result is shown in Fig. 3(b), where the broadband CPA response is obvious. Coherent absorption with large values $\geq 90\%$ is obtained over a broad frequency range, from 2 THz to 3.5 THz, interestingly by using an extremely subwavelength thin structure. The broadband coherent absorption response is due to the hybridization and coupling of the narrowband resonant CPA responses of the front and back graphene metasurfaces with different side lengths $L_1$ and $L_2$, respectively. Thus, asymmetry in the structure is required to achieve a broadband CPA response since a symmetric bifacial graphene metasurface design will only exhibit CPA at a single resonant frequency point.

Tunability is a unique feature to graphene that enables dynamical control in the coherent absorption of the proposed device by controlling its Fermi level. In order to investigate the Fermi level's effect on the CPA performance, we plot the distributions of the output coefficient when $E_F$ varies from 0.05 to 0.95 eV and the incident frequency from both sides is swept from 1 to 5 THz. The result is shown in Fig. 4(a). Note that the Drude expression of graphene conductivity still works in this THz frequency range, where interband transitions are absent. Furthermore, we keep using the same $\alpha$ complex value $\alpha = 0.9 e^{i*10°}$ utilized to produce the results shown in Fig. 3(b), which defines the relation between the two counter-propagating incident waves. It is demonstrated in Fig. 4(a) that the Fermi level has a large impact on the output coefficient value. First, a substantial blue shift in the resonant frequency is observed as the Fermi level increases. This is in agreement with the general relationship between resonant frequency and Fermi level in graphene patches that can simply be expressed as $f_r \propto \sqrt{E_F / L}$ [22,63], where $L$ is the side length of each graphene patch. Second, the Fermi level variation can alter the output coefficient at the resonance because it effectively changes graphene's properties. Finally, we derive from Fig. 4(a) that broadband CPA can be achieved between 0.3 to 0.5 eV, which are moderate doping level values. Hence, impractically high doping graphene values are not required for the current tunable and broadband CPA design. As a result, it is not required to increase the applied gate voltage, which is proportional to the Fermi level values, significantly to obtain the presented broadband CPA performance.

Next, we analyse the graphene's DC mobility effect on the proposed asymmetric metasurface's CPA response. This feature characterizes the electronic quality of graphene and determines the phenomenological scattering rate $\Gamma$ that can be affected by many factors, such as temperature, Fermi level and graphene sample quality [64]. Figure 4(b) shows the output coefficient of the proposed device as a function of DC mobility and incident frequency. We choose to vary the DC mobility from 0.05 to 1 m$^2$/V$s$ even though it has been reported that mobility values higher than $20\,m^2/Vs$ are possible when extrinsic disorder in graphene's exfoliation process is eliminated [65]. We keep the Fermi level constant and equal to $E_F = 0.4\,eV$ in the results presented in Fig. 4(b). It is obvious that the CPA resonance is almost constant when we vary the mobility from 0.05 to $1\,m^2/Vs$. Moreover, the output coefficient decreases to zero for some frequencies when the mobility increases to $0.4\,m^2/Vs$ and beyond. Hence, CPA can always be obtained with mobility values higher than $0.4\,m^2/Vs$. These relative low mobility values can be easily implemented by using conventional graphene exfoliation approaches. Hence, no special requirements are expected during the potential fabrication process of the proposed bifacial graphene metasurface. By inspecting the results in Fig. 4, we conclude that the most efficient way to modulate the CPA response of the proposed structure is to tune graphene's Fermi level.

According to the theoretical analysis presented in section 2, the ratio value $\alpha$ of the two counter-propagating input waves plays an important role in realizing the presented broadband CPA performance. Figure 5 confirms this conclusion and, interestingly, demonstrates that this ratio can control the CPA response in a nonlinear way due to the current structurally-asymmetric metasurface design. Hence, we calculate the output coefficient as a function of frequency and the amplitude ratio $|\alpha|$ and the result is shown in Fig. 5(a). The phase difference between the two counter-propagating waves is fixed to $10°$ at this plot, similar to the value used in Fig. 3(b). Coherent absorption values $\geq 90\%$ are obtained over a broad frequency range from 2.1 to 3.2 THz when the amplitude ratio is larger than the value 0.48. By using smaller amplitude ratios, greater difference between the two input waves exists, leading to weaker destructive interference. As a result, it is impossible to achieve CPA with small amplitude ratios. However, the small amplitude regime will be used to achieve the inverse counterpart of absorption, i.e., amplification, as it will be demonstrated later in this paper.

The effect of phase difference $\varphi$ is further investigated in Fig. 5(b). This figure illustrates the output coefficient as a function of the amplitude ratio $|\alpha|$ under dissimilar phase difference values $\varphi$. In this case, the frequency of the two incident waves is fixed to 2.4 THz. The CPA performance is evident when the amplitude of the ratio and the phase difference of the incident waves are $|\alpha|=0.9$ and $\varphi=10°$, respectively. Note that the $\varphi=0°$ case is also plotted in Fig. 5(b) and is found to have slightly higher output coefficient values compared to the $\varphi=10°$ scenario. The other interesting point is that perfect transmission can be realized with the same device by using dissimilar phase difference $\varphi$ values. For example, the output coefficient can reach to 1 when the phase difference is $\varphi=180°$ and the amplitude ratio is $|\alpha|=1.1$, which is clearly depicted in Fig. 5(b). Interestingly, the proposed device can be tuned in a nonlinear fashion, as illustrated in Fig. 5(b), by varying both the amplitude and phase of the parameter $\alpha$ computed by the ratio of the two counter-propagating input waves. This nonlinear tunable response is caused by purely light interference effects (light-light interactions [44]) and not due to the optical nonlinearity of the used materials, which is usually very weak and requires high input intensities to be excited. Note that the presented nonlinear response is a direct consequence of the proposed structure's geometrical asymmetry and cannot be achieved by symmetric configurations.

When two different incident waves $E_{in1}$ and $E_{in2}$ illuminate the proposed device from both sides, asymmetric devices can reach and operate in the small-signal amplificationregime, which is analogous to the field effect transistor (FET) amplifier operation but now working for light (all-optical operation) [66]. In this scenario, the intensity $I_{in2}$ of the incident wave $E_{in2}$ from the front side is fixed and the intensity $I_{in1}$ in the other side plays the role of the small input signal, acting as a probe, which will be used to amplify the total output response. The output signal intensity $I_{out1}$ from the back side is computed to demonstrate the amplifying performance of the presented asymmetric device. The gain coefficient $G$ of the proposed amplifier is defined to demonstrate the amplification response in a quantitative way. It is computed by the derivative of $I_{out1}$ with respect to $I_{in1}$ and is expressed as: $G = dI_{out1}/dI_{in1}$ [44]. This formula is combined with Eq. (2) and $\alpha = E_{in1}/E_{in2}$, leading to the final gain coefficient equation:

$$G = |r_1|^2 + \text{Re}\{r_1 t^* e^{i\varphi}\}/|\alpha| \tag{5}$$

By inspecting the derived Eq. (5), it can be concluded that the gain coefficient $G$ can become infinitely large if $\alpha$ becomes infinitesimal small. This conclusion is consistent with the results obtained by full-wave simulations and shown in Fig. 6(a), where the gain coefficient is computed and plotted as a function of the amplitude ratio between the two input waves. In this case, the phase difference between the two incident waves is fixed to $\varphi = 180°$, where perfect constructive interference between the two incident waves is achieved leading to amplification, which is the inverse response compared to absorption that was caused due to destructive interference, as was shown before. The used frequency is $f$ = 2.4 THz, where the maximum output coefficient is obtained, as it was demonstrated before in Fig. 5(b). However, we need to emphasize that the system will have no gain if $\alpha$ is reduced to zero. It is worth noting that the gain can always be larger than 1 only when $0 < |\alpha| \leq 0.3$, as verified by the zoomed-in inset plot in Fig. 6(a). Importantly, the phase response, defined as the phase difference between the input $E_{in1}$ and output $E_{out1}$ waves, is kept constant when the amplitude ratio $|\alpha|$ varies, as shown in Fig. 6(a) (dashed line), implying that the amplified output wave is always in-phase with the input signal, a very important functionality to achieve a successful amplification process.

In order to further investigate the broadband nature of the obtained gain coefficient of the presented four-port all-optical THz amplification device, we calculate the gain response as a function of frequency. The result is presented in Fig. 6(b), where the ratio between the two input waves is fixed to $\alpha = 0.1 e^{i*180°}$. The gain coefficient keeps its highest value (equal to 3) over a broad frequency range from 2.2 to 3.1 THz. Moreover, the gain can always be larger than one in an even broader frequency range from 1.3 THz to 5.9 THz. Hence, the THz signal can be amplified within the broad bandwidth of 4.6 THz, indicating that the proposed device can work as an efficient, broadband and ultrathin all-optical amplifier. Finally, it is important to mention that the presented broadband amplifying response is not possible to be achieved by using symmetric CPA devices.

## 4. Conclusion

Summarizing, we have investigated and shown tunable and broadband CPA response based on an asymmetric bifacial graphene metasurface design with ultrathin subwavelength dimensions. We have also demonstrated the nonlinear response of this device under illuminations with different intensities from both front and back sides. Thanks to the asymmetric light interference, the reflection spectra is different from opposite sides of the proposed device in the case of single wave illumination. This effect happens over a broad frequency range from 2 THz to 3.5 THz and leads to $\geq 90\%$ broadband coherent absorption when two counter-propagating incident waves are considered. By dynamically controlling the Fermi level and the DC mobility of graphene, tunable broadband CPA response is demonstrated. In addition, and more importantly, nonlinear CPA response and amplification performance are obtained based on the proposed ultrathin asymmetric design. Almost perfect absorption is attained over a wide frequency range with the amplitude of the two incident waves ratio being larger than 0.48. However, smaller amplitude ratios lead to the presented strong amplification performance, consisting the inverse operation compared to absorption. These interesting effects are based on the nonlinear response of the proposed structure due to strong interference between the incident waves (light-light interactions) and not the usually weak nonlinear material properties. The designed asymmetric bifacial graphene metasurfaces can amplify the signal in an efficient way and over a broad frequency range. Potential applications of the proposed ultrathin compact THz device are modulators, sensors, coherent photodetectors, and all-optical small-signal amplifiers.

## Acknowledgements

This work was partially supported by the National Science Foundation Nebraska Materials Research Science and Engineering Center (Grant No. DMR1420645), Office of Naval Research Young Investigator Program (ONR-YIP) Award (Grant No. N00014-19-1- 2384) and National Science Foundation Nebraska-EPSCoR (Grant No. OIA1557417).

# References

1. M. I. Katsnelson, *Graphene* (Cambridge University Press, 2012).
2. F. Bonaccorso, Z. Sun, T. Hasan, and A. C. Ferrari, "Graphene photonics and optoelectronics," Nat. Photonics **4**(9), 611–622 (2010).
3. H. Yan, X. Li, B. Chandra, G. Tulevski, Y. Wu, M. Freitag, W. Zhu, P. Avouris, and F. Xia, "Tunable infrared plasmonic devices using graphene/insulator stacks," Nat. Nanotechnol. **7**(5), 330–334 (2012).
4. P.-Y. Chen, C. Argyropoulos, M. Farhat, and J. S. Gomez-Diaz, "Flatland plasmonics and nanophotonics based on graphene and beyond," Nanophotonics **6**(6), 1239–1262 (2017).
5. K. F. Mak, M. Y. Sfeir, Y. Wu, C. H. Lui, J. A. Misewich, and T. F. Heinz, "Measurement of the Optical Conductivity of Graphene," Phys. Rev. Lett. **101**(19), 196405 (2008).
6. F. Liu, Y. D. Chong, S. Adam, and M. Polini, "Gate-tunable coherent perfect absorption of terahertz radiation in graphene," 2D Mater. **1**(3), 031001 (2014).
7. B. Jin, T. Guo, and C. Argyropoulos, "Enhanced third harmonic generation with graphene metasurfaces," J. Opt. **19**(9), 094005 (2017).
8. T. Guo, B. Jin, and C. Argyropoulos, "Hybrid Graphene-Plasmonic Gratings to Achieve Enhanced Nonlinear Effects at Terahertz Frequencies," Phys. Rev. Appl. **11**(2), 024050 (2019).
9. C. Argyropoulos, "Enhanced transmission modulation based on dielectric metasurfaces loaded with graphene," Opt. Express **23**(18), 23787 (2015).
10. S. Thongrattanasiri, F. H. L. Koppens, and F. J. García de Abajo, "Complete Optical Absorption in Periodically Patterned Graphene," Phys. Rev. Lett. **108**(4), 047401 (2012).
11. F. H. L. Koppens, D. E. Chang, and F. J. García de Abajo, "Graphene Plasmonics: A Platform for Strong Light–Matter Interactions," Nano Lett. **11**(8), 3370–3377 (2011).
12. Y. M. Qing, H. F. Ma, S. Yu, and T. J. Cui, "Tunable dual-band perfect metamaterial absorber based on a graphene-SiC hybrid system by multiple resonance modes," J. Phys. D. Appl. Phys. **52**(1), 015104 (2019).
13. Y. M. Qing, H. F. Ma, and T. J. Cui, "Flexible control of light trapping and localization in a hybrid Tamm plasmonic system," Opt. Lett. **44**(13), 3302 (2019).
14. Y. M. Qing, H. F. Ma, Y. Z. Ren, S. Yu, and T. J. Cui, "Near-infrared absorption-induced switching effect via guided mode resonances in a graphene-based metamaterial," Opt. Express **27**(4), 5253 (2019).
15. B.-X. Wang, "Quad-Band Terahertz Metamaterial Absorber Based on the Combining of the Dipole and Quadrupole Resonances of Two SRRs," IEEE J. Sel. Top. Quantum Electron. **23**(4), 1–7 (2017).
16. M. Jablan, H. Buljan, and M. Soljačić, "Plasmonics in graphene at infrared frequencies," Phys. Rev. B **80**(24), 245435 (2009).
17. Z. Li, K. Yao, F. Xia, S. Shen, J. Tian, and Y. Liu, "Graphene Plasmonic Metasurfaces to Steer Infrared Light," Sci. Rep. **5**(1), 12423 (2015).
18. Y. M. Qing, H. F. Ma, and T. J. Cui, "Theoretical Analysis of Tunable Multimode Coupling in a Grating-Assisted Double-Layer Graphene Plasmonic System," ACS Photonics **6**(11), 2884–2893 (2019).

19. Y. M. Qing, H. F. Ma, and T. J. Cui, "Investigation of strong multimode interaction in a graphene-based hybrid coupled plasmonic system," Carbon N. Y. **145**, 596–602 (2019).
20. P.-Y. Chen, C. Argyropoulos, and A. Alu, "Terahertz Antenna Phase Shifters Using Integrally-Gated Graphene Transmission-Lines," IEEE Trans. Antennas Propag. **61**(4), 1528–1537 (2013).
21. P. Chen and A. Alù, "Atomically Thin Surface Cloak Using Graphene Monolayers," ACS Nano **5**(7), 5855–5863 (2011).
22. T. Guo and C. Argyropoulos, "Broadband polarizers based on graphene metasurfaces," Opt. Lett. **41**(23), 5592 (2016).
23. B.-X. Wang, G.-Z. Wang, and T. Sang, "Simple design of novel triple-band terahertz metamaterial absorber for sensing application," J. Phys. D. Appl. Phys. **49**(16), 165307 (2016).
24. B.-X. Wang, G.-Z. Wang, and L.-L. Wang, "Design of a Novel Dual-Band Terahertz Metamaterial Absorber," Plasmonics **11**(2), 523–530 (2016).
25. B.-X. Wang, Y. He, P. Lou, and W. Xing, "Design of a dual-band terahertz metamaterial absorber using two identical square patches for sensing application," Nanoscale Adv. **2**(2), 763–769 (2020).
26. N. K. Emani, T.-F. Chung, X. Ni, A. V. Kildishev, Y. P. Chen, and A. Boltasseva, "Electrically Tunable Damping of Plasmonic Resonances with Graphene," Nano Lett. **12**(10), 5202–5206 (2012).
27. Z. Fang, Y. Wang, A. E. Schlather, Z. Liu, P. M. Ajayan, F. J. García de Abajo, P. Nordlander, X. Zhu, and N. J. Halas, "Active Tunable Absorption Enhancement with Graphene Nanodisk Arrays," Nano Lett. **14**(1), 299–304 (2014).
28. W. Wan, Y. Chong, L. Ge, H. Noh, A. D. Stone, and H. Cao, "Time-Reversed Lasing and Interferometric Control of Absorption," Science (80 ). **331**(6019), 889–892 (2011).
29. Y. D. Chong, L. Ge, H. Cao, and A. D. Stone, "Coherent Perfect Absorbers: Time-Reversed Lasers," Phys. Rev. Lett. **105**(5), 053901 (2010).
30. C. F. Gmachl, "Suckers for light," Nature **467**(7311), 37–39 (2010).
31. D. G. Baranov, A. Krasnok, T. Shegai, A. Alù, and Y. Chong, "Coherent perfect absorbers: linear control of light with light," Nat. Rev. Mater. **2**(12), 17064 (2017).
32. Y. Li and C. Argyropoulos, "Tunable nonlinear coherent perfect absorption with epsilon-near-zero plasmonic waveguides," Opt. Lett. **43**(8), 1806 (2018).
33. Y. Sun, W. Tan, H. Li, J. Li, and H. Chen, "Experimental Demonstration of a Coherent Perfect Absorber with PT Phase Transition," Phys. Rev. Lett. **112**(14), 143903 (2014).
34. X. Feng, J. Zou, W. Xu, Z. Zhu, X. Yuan, J. Zhang, and S. Qin, "Coherent perfect absorption and asymmetric interferometric light-light control in graphene with resonant dielectric nanostructures," Opt. Express **26**(22), 29183 (2018).
35. T. Guo and C. Argyropoulos, "Tunable and broadband coherent perfect absorption by ultrathin black phosphorus metasurfaces," J. Opt. Soc. Am. B **36**(11), 2962 (2019).
36. M. Pu, Q. Feng, M. Wang, C. Hu, C. Huang, X. Ma, Z. Zhao, C. Wang, and X. Luo, "Ultrathin broadband

nearly perfect absorber with symmetrical coherent illumination," Opt. Express **20**(3), 2246 (2012).
37. G. Pirruccio, L. Martín Moreno, G. Lozano, and J. Gómez Rivas, "Coherent and Broadband Enhanced Optical Absorption in Graphene," ACS Nano **7**(6), 4810–4817 (2013).
38. X. Hu and J. Wang, "High-speed gate-tunable terahertz coherent perfect absorption using a split-ring graphene," Opt. Lett. **40**(23), 5538 (2015).
39. V. Thareja, J.-H. Kang, H. Yuan, K. M. Milaninia, H. Y. Hwang, Y. Cui, P. G. Kik, and M. L. Brongersma, "Electrically Tunable Coherent Optical Absorption in Graphene with Ion Gel," Nano Lett. **15**(3), 1570–1576 (2015).
40. Y. Fan, Z. Liu, F. Zhang, Q. Zhao, Z. Wei, Q. Fu, J. Li, C. Gu, and H. Li, "Tunable mid-infrared coherent perfect absorption in a graphene meta-surface," Sci. Rep. **5**(1), 13956 (2015).
41. C. Argyropoulos and B. Jin, "Nonlinear graphene metasurfaces with advanced electromagnetic functionalities," in *Plasmonics: Design, Materials, Fabrication, Characterization, and Applications XVI*, T. Tanaka and D. P. Tsai, eds. (SPIE, 2018), p. 62.
42. N. Volet, A. Spott, E. J. Stanton, M. L. Davenport, L. Chang, J. D. Peters, T. C. Briles, I. Vurgaftman, J. R. Meyer, and J. E. Bowers, "Semiconductor optical amplifiers at 2.0-μm wavelength on silicon," Laser Photon. Rev. **11**(2), 1600165 (2017).
43. T. Guo, L. Zhu, P.-Y. Chen, and C. Argyropoulos, "Tunable terahertz amplification based on photoexcited active graphene hyperbolic metamaterials [Invited]," Opt. Mater. Express **8**(12), 3941 (2018).
44. X. Fang, K. F. MacDonald, and N. I. Zheludev, "Controlling light with light using coherent metadevices: all-optical transistor, summator and invertor," Light Sci. Appl. **4**(5), e292–e292 (2015).
45. Pai-Yen Chen, Haiyu Huang, D. Akinwande, and A. Alù, "Distributed Amplifiers Based on Spindt-Type Field-Emission Nanotriodes," IEEE Trans. Nanotechnol. **11**(6), 1201–1211 (2012).
46. Z. J. Wong, Y.-L. Xu, J. Kim, K. O'Brien, Y. Wang, L. Feng, and X. Zhang, "Lasing and anti-lasing in a single cavity," Nat. Photonics **10**(12), 796–801 (2016).
47. S. Longhi, "PT-symmetric laser absorber," Phys. Rev. A **82**(3), 031801 (2010).
48. K. S. Novoselov, V. I. Fal′ko, L. Colombo, P. R. Gellert, M. G. Schwab, and K. Kim, "A roadmap for graphene," Nature **490**(7419), 192–200 (2012).
49. Z. Lu and W. Zhao, "Nanoscale electro-optic modulators based on graphene-slot waveguides," J. Opt. Soc. Am. B **29**(6), 1490 (2012).
50. G. W. Hanson, "Dyadic Green's Functions for an Anisotropic, Non-Local Model of Biased Graphene," IEEE Trans. Antennas Propag. **56**(3), 747–757 (2008).
51. H. Nasari and M. S. Abrishamian, "All-optical tunable notch filter by use of Kerr nonlinearity in the graphene microribbon array," J. Opt. Soc. Am. B **31**(7), 1691 (2014).
52. J. Horng, C.-F. Chen, B. Geng, C. Girit, Y. Zhang, Z. Hao, H. A. Bechtel, M. Martin, A. Zettl, M. F. Crommie, Y. R. Shen, and F. Wang, "Drude conductivity of Dirac fermions in graphene," Phys. Rev. B **83**(16), 165113 (2011).

53. K. S. Novoselov, "Electric Field Effect in Atomically Thin Carbon Films," Science (80). **306**(5696), 666–669 (2004).
54. COMSOL Multiphysics, http://www.comsol.com/ .
55. X. Li, C. W. Magnuson, A. Venugopal, R. M. Tromp, J. B. Hannon, E. M. Vogel, L. Colombo, and R. S. Ruoff, "Large-Area Graphene Single Crystals Grown by Low-Pressure Chemical Vapor Deposition of Methane on Copper," J. Am. Chem. Soc. **133**(9), 2816–2819 (2011).
56. M. C. Lemme, D. C. Bell, J. R. Williams, L. A. Stern, B. W. H. Baugher, P. Jarillo-Herrero, and C. M. Marcus, "Etching of Graphene Devices with a Helium Ion Beam," ACS Nano **3**(9), 2674–2676 (2009).
57. Z. Fang, S. Thongrattanasiri, A. Schlather, Z. Liu, L. Ma, Y. Wang, P. M. Ajayan, P. Nordlander, N. J. Halas, and F. J. García de Abajo, "Gated Tunability and Hybridization of Localized Plasmons in Nanostructured Graphene," ACS Nano **7**(3), 2388–2395 (2013).
58. C. Wang, W. Liu, Z. Li, H. Cheng, Z. Li, S. Chen, and J. Tian, "Dynamically Tunable Deep Subwavelength High-Order Anomalous Reflection Using Graphene Metasurfaces," Adv. Opt. Mater. **6**(3), 1701047 (2018).
59. J. Kim, H. Son, D. J. Cho, B. Geng, W. Regan, S. Shi, K. Kim, A. Zettl, Y.-R. Shen, and F. Wang, "Electrical Control of Optical Plasmon Resonance with Graphene," Nano Lett. **12**(11), 5598–5602 (2012).
60. F. Monticone, C. A. Valagiannopoulos, and A. Alù, "Parity-Time Symmetric Nonlocal Metasurfaces: All-Angle Negative Refraction and Volumetric Imaging," Phys. Rev. X **6**(4), 041018 (2016).
61. F. Xia, T. Mueller, Y. Lin, A. Valdes-Garcia, and P. Avouris, "Ultrafast graphene photodetector," Nat. Nanotechnol. **4**(12), 839–843 (2009).
62. M. Liu, X. Yin, E. Ulin-Avila, B. Geng, T. Zentgraf, L. Ju, F. Wang, and X. Zhang, "A graphene-based broadband optical modulator," Nature **474**(7349), 64–67 (2011).
63. Z. H. Zhu, C. C. Guo, K. Liu, J. F. Zhang, W. M. Ye, X. D. Yuan, and S. Q. Qin, "Electrically tunable polarizer based on anisotropic absorption of graphene ribbons," Appl. Phys. A **114**(4), 1017–1021 (2014).
64. J.-H. Chen, C. Jang, S. Xiao, M. Ishigami, and M. S. Fuhrer, "Intrinsic and extrinsic performance limits of graphene devices on SiO2," Nat. Nanotechnol. **3**(4), 206–209 (2008).
65. S. V. Morozov, K. S. Novoselov, M. I. Katsnelson, F. Schedin, D. C. Elias, J. A. Jaszczak, and A. K. Geim, "Giant Intrinsic Carrier Mobilities in Graphene and Its Bilayer," Phys. Rev. Lett. **100**(1), 016602 (2008).
66. M. A. Andersson, O. Habibpour, J. Vukusic, and J. Stake, "10 dB small-signal graphene FET amplifier," Electron. Lett. **48**(14), 861 (2012).

**Figures**

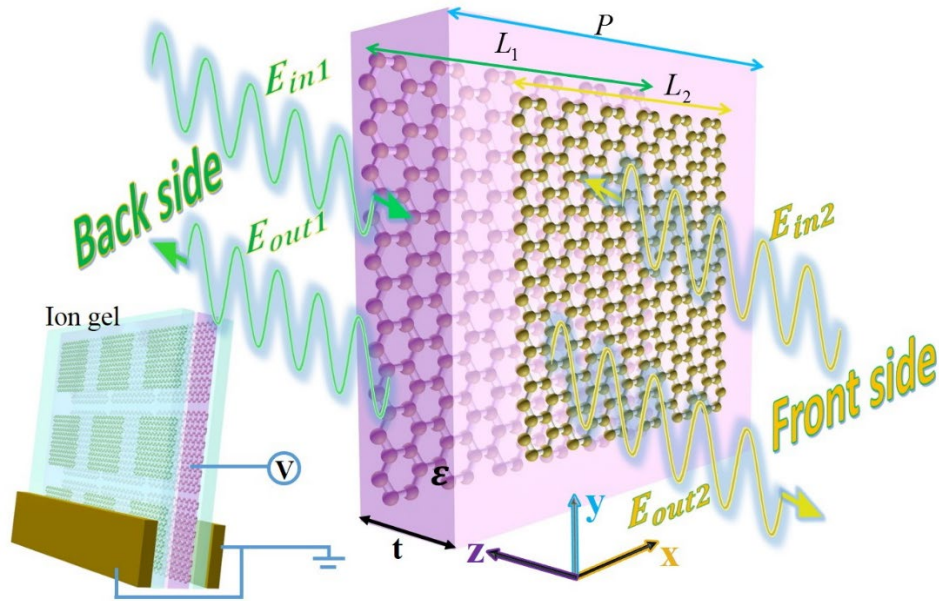

Figure 1. Unit cell design of the asymmetric bifacial metasurface made of different size periodic graphene patches. The inset presents the schematic of an ion gel-gating approach to dope the graphene and tune the asymmetric bifacial graphene metasurface response.

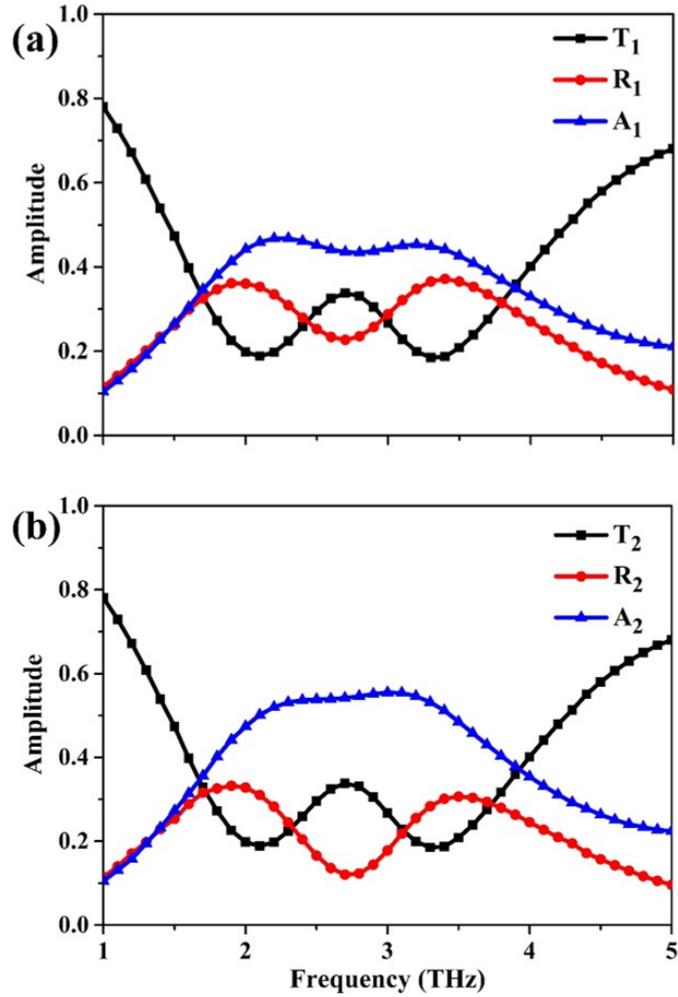

Figure 2. Transmission, reflection and absorption coefficient spectra of the proposed structure excited by a normal incident TM-polarized wave impinging from the (a) back ($E_{in1}$) and (b) front ($E_{in2}$) directions.

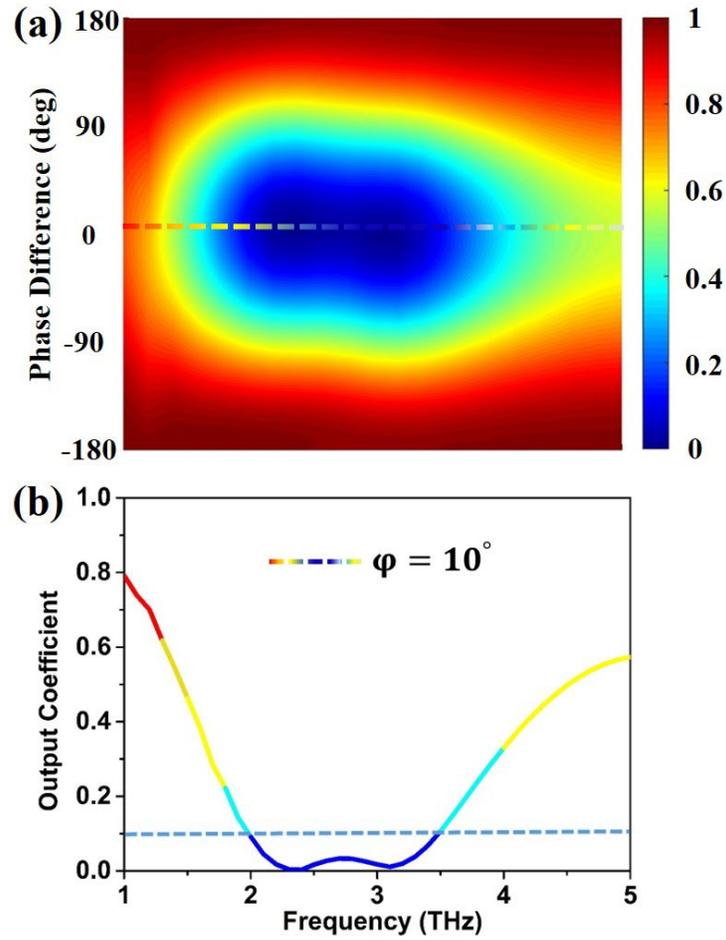

Figure 3. (a) Output coefficient of the proposed device as a function of frequency and the phase difference of the two incident waves. The minimum output coefficient is obtained at the frequency of 2.4 THz when using a phase difference of φ = 10°. (b) Output coefficient as a function of frequency under phase difference φ = 10°. This result is the cross-section plot following the dashed line in caption (a).

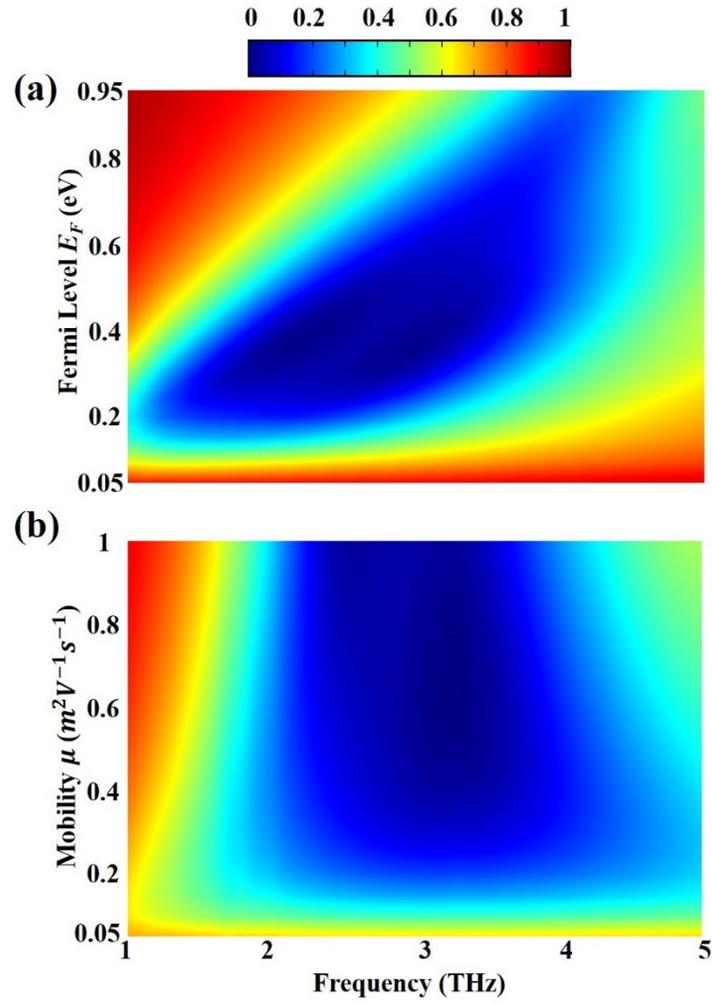

Figure 4. Output coefficient of the proposed device as a function of frequency and (a) Fermi level or (b) DC mobility of graphene.

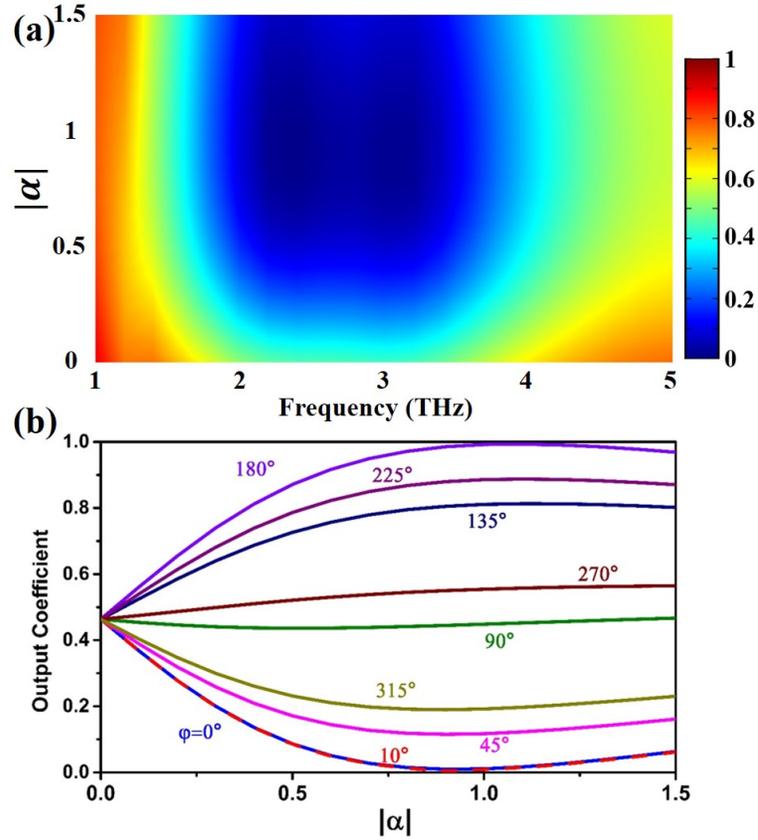

Figure 5. (a) Output coefficient of the proposed device as a function of frequency and amplitude ratio |α|. The phase difference is fixed to φ = 10°. (b) Dependence of output coefficient on the incident wave ratio | α | and the phase difference φ. Nonlinear CPA response is achieved due to the asymmetric geometry of the proposed device.

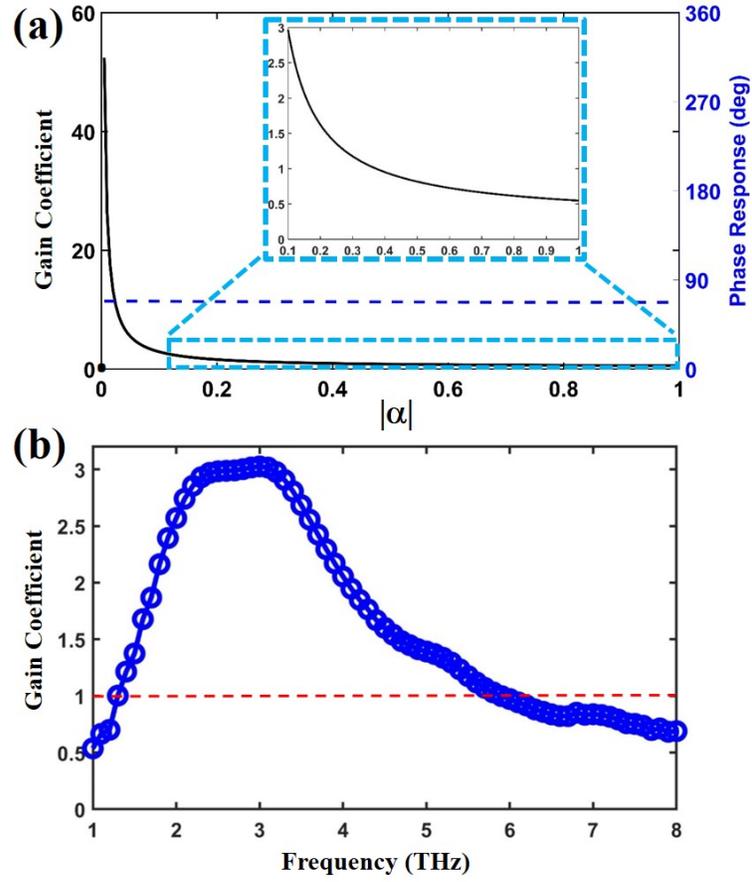

Figure 6. (a) Gain coefficient and phase response between input and output waves as a function of the asymmetric amplitude ratio $|\alpha|$. (b) Gain coefficient as a function of frequency by using a fixed ratio between the two input waves: $\alpha = 0.1e^{i*180°}$.

21